\magnification\magstep1
\centerline{\bf Alpha and Electro-weak Coupling}
\centerline{by}
\centerline{James G. Gilson, Mathematics QMW London}
\centerline{PIRT Conference September 2000 Imperial College London}
\vskip .3cm

\vskip .5cm
\centerline{\bf Abstract}

It is shown that the fine structure constant $\alpha$ has the same value that {\it characterises\/} a relation, denoted by $\alpha _{137}(29\times 137)$, between a representation of the cyclic group of order $29\times 137$ and the induced representation for the cyclic subgroup of order $137$. The value of this characteristic is $\alpha _{137}(29\times 137) =0.007297352532...$. The complementary characteristic $\alpha _{29}(29\times 137)=0.034280626357...$\ \ for the cyclic subgroup of order $29$ is shown to represent the gauge theory electro-weak coupling quantity $g^2/4\pi$. Kinematic aspects of the representation geometry are discussed and a generalized version of the Weinberg electro-weak mixing angle is introduced.  
\vskip .5cm
\centerline{\bf 1 Euclidean Aspect}
\vskip .5cm 
In more detail, this result can be explained in terms of simple Euclidean geometry as follows. 
Consider the special case of a cyclic group $G_{n_1n_2}$ of order $n_1n_2$ with only the two prime subgroups $G_{n_1}$ and $G_{n_2}$ of orders prime numbers $n_1$ and $n_2$ respectively. The group $G_{n_1n_2}$ is isomorphic with the fixed plane symmetry rotations of an equilateral $n_1n_2$ sided polygon $P_{n_1n_2}$ with rotation angle unit $2\chi _{n_1n_2}=2\pi /{n_1n_2}$ subtended by a side at its centre. It is convenient to express these polygonal angles in terms of their half values such as $\chi _{n_1n_2}$. With the polygon $P_{n_1n_2}$ we can associate a collection of seven geometrical objects, concentric circles and lesser-sided polygons. There are the inscribed and circumscribed circles of $P_{n_1n_2}$, $(C_{n_1n_2,b},\quad C_{n_1n_2,a})$ and a circle $C_{n_1n_2,c}$ of perimeter equal in length to that of $P_{n_1n_2}$ which threads through the sides of $P_{n_1n_2}$. The two subgroups $G_{n_1}$ and $G_{n_2}$ have their own associated cyclical symmetry polygons $P_{n_1}$ with angular steps $2\chi _{n_1}=2\pi/n_1$ and $P_{n_2}$ with angular steps $2\chi _{n_2}=2\pi/n_2$. These polygons enclose the polygon $P_{n_1n_2}$ touching it along common sides with the greater sided polygon being inside the lesser sided one. These last two polygons have a common inscribed circle $C_{n_1n_2,b}$ with $P_{n_1n_2}$. This circle has already been counted but they have the separate circumscribed circles $(C_{n_1,d},C_{n_2,e})$ to be added to the collection which including the initial polygon contains eight concentric geometrical objects. Let us suppose that $n_2\le n_1$ then the lesser-sided polygon $P_{n_2}$ will be outside $P_{n_1}$.

Let us designate a special direction radial line as a line through the centre of the system of concentric polygons which also passes through a mid point of the sides of both the sub-polygons $P_{n_1}$ and $P_{n_2}$.  
Projecting a radius $r_{n_1,d}$ of the circumscribed circle $C_{n_1,d}$ of $P_{n_1}$ through the half angle $\chi _{n_1}=\pi /n_1$ on to the special direction gives the radius $r_{n_1n_2,b}$, of the inscribed circle $C_{n_1n_2,b}$.
$$ r_{n_1,d} \cos (\chi _{n_1})= r_{n_1n_2,b}.$$
Similarly for the case of the other subgroup
$$ r_{n_2,e} \cos (\chi _{n_2})= r_{n_1n_2,b}.$$
However, If the slightly larger radius of the threading circle $r_{n_1n_2,c}$ is projected from a radius $ r_{n_1,d} $ of the circumscribed circle $C_{n_1,d}$ of $P_{n_1}$, it will have to have come from a slightly smaller half angle $\chi _{n_1}^*\le\pi/n_1$, say. Thus the numerical value of the angle $\chi _{n_1}^*$ is given by
$$ r_{n_1,d} \cos (\chi _{n_1}^*)= r_{n_1n_2,c}.$$
Similarly for the case of the other subgroup $G_{n_2}$
$$ r_{n_2,e} \cos (\chi _{n_2}^*)= r_{n_1n_2,c}.$$

We note that the length of the circumference of the threading circle $C_{n_1n_2,c}$ is $C=2n_1n_2 \tan (\pi/(n_1n_2)) r_{n_1n_2,b}$, by definition the same as the perimeter length of the polygon $P_{n_1n_2}$. This implies that the radius of the threading circle can be expressed in the form $r_{n_1n_2,c}=C/{2\pi}= n_1n_2 r_{n_1n_2,b} \tan (\pi/(n_1n_2))/\pi.$
This in turn implies that
$$r_{n_1n_2,c}= r_{n_1,d} n_1n_2\cos (\chi _{n_1})\tan (\pi/(n_1n_2))/\pi=r_{n_1,d}\cos (\chi ^*_{n_1}).$$
Consequently
$$\cos (\chi ^*_{n_1})/n_1=n_2\cos (\chi_{n_1})\tan (\pi /(n_1n_2))/\pi.$$
Using the other subgroup $G_{n_2}$ 
$$r_{n_1n_2,c}= r_{n_2,e} n_1n_2\cos (\chi _{n_2})\tan (\pi/(n_1n_2))/\pi =r_{n_2,e}\cos (\chi ^*_{n_2}).$$
Consequently
$$\cos (\chi ^*_{n_2})/n_2=n_1\cos (\chi_{n_2})\tan (\pi /(n_1n_2))/\pi.$$

There is no quantum theory being used in this analysis but we shall use the terminology of quantum theory {\it analogously\/} for quantities that can naturally be express as some integer times some special unit of the quantity involved. For example if $l=nl_c$ where $n$ is an integer and $l_c$ is some special unit of length, the length $l$ will be said to be quantized with quantum number or eigenvalue $n$ and quantum of length $l_c$. Consequently, the radius $r_{n_1n_2,c}$ can be expressed as a quantized length in the two ways $r_{n_1n_2,c}=n_1l_{n_1}=n_2l_{n_2}$ in terms of quantum lengths
$$l_{n_1}=r_{n_1,d}n_2\cos (\chi _{n_1})\tan (\pi/(n_1n_2))/\pi$$ or $$l_{n_2}=r_{n_2,e}n_1\cos (\chi _{n_2})\tan (\pi/(n_1n_2))/\pi$$ with eigenvalues $n_2$ and $n_1$ respectively according as which subgroup $G_{n_1}$ or $G_{n_2}$ is taken to be the generator.

The ratios of the quantum lengths $l_{n_1},l_{n_2}$ to their generating subgroup polygon radii $r_{n_1,d}$ and $r_{n_2,e}$ express a characteristic measure of their relation with the $G_{n_1n_2}$ group structure. Let us denote the characteristic values by $\alpha _{n_1} (n_1n_2) $ and $\alpha _{n_2}(n_1n_2) $ so that we have
$$\alpha _{n_1}(n_1n_2) =\cos (\chi ^*_{n_1})/n_1 = n_2\cos (\chi _{n_1})\tan (\pi/(n_1n_2))/\pi$$
and
$$\alpha _{n_2}(n_1n_2) =\cos (\chi ^*_{n_2})/n_2= n_1\cos (\chi _{n_2})\tan (\pi/(n_1n_2))/\pi.$$
For the special case $ n_1=137,\  n_2=29$ the values of these two characteristics are
$$\alpha _{137}(29\times 137)=0.007297352532$$
and
$$ \alpha _{29}(29\times 137) =0.034280626357$$
both evaluated to twelve decimal places.
The experimental value for the fine structure constant according to the latest measurements is
$$\alpha=0.007297352533(27).$$
The $27$ in brackets is a measurement uncertainty of $\pm 27$ for the last two decimal places. 
The characteristic value $\alpha _{29}(29\times 137)$ is identifiable with the numerical value of the {\it theoretical\/} gauge theory electro-weak coupling constant $g^2/{4\pi}=\alpha /\sin ^2 (\theta _W)$. The theoretical value assumed by this quantity depends on the value that is given to the Weinberg weak mixing angle $\theta _W$. If we take the value from the  Georgi-Glashow model$^1$ where $\sin ^2 (\theta _W)=0.21$,
$g^2/{4\pi}=0.035$. If we take the Weinberg value $\sin ^2 (\theta _W)=0.203$ then $g^2/{4\pi}=0.036$. Section 21 of Weinberg's second volume$^2$ "The Quantum Theory of Fields" gives values and remarks on this issue. An accurate measured value for $g^2/4\pi$ can be obtained by using the information about the W-boson-lepton vertex interaction which is controlled by the coupling constant $\alpha _W=g^2_W/4\pi= (g^2/4\pi)/8$ and which is related to the accurately known Firmi coupling constant $G_F$. This gives the formula and value,
$$g^2/4\pi =\sqrt 2 M_W^2 G_F/\pi=0.033882$$
when the values used for the $M_W$ boson mass and the Fermi coupling constant $G_F$ are taken to be $M_W=80.33\  GeV/c^2$ and $G_F=1.16639\times 10^{-5}\  GeV^{-2}$ respectively. This measurement value differs from the value above predicted by the present theory using the identification $g^2/4\pi =\alpha _{29}(29\times 137)$ by $4$ parts in $10^4$ parts. This is a good agreement between theory and measurement for quantities in this extra-QED context, but clearly not of the same order of exactitude as the value predicted for the fine structure constant$^{3,7}$.
Aspects of the W-boson-lepton vertex evaluation can be found on page 196 of the book$^5$ "Particle Physics" by B.R.Martin \& G. Shaw. Except for the highly speculative and uncertain supersymmetry theoretical constructs, all the current theories of the week interaction coupling values for $\sin ^2 (\theta _W)$ are {\it out\/} of agreement at the second decimal place with what is considered to be a very accurate physical measurement value $\sin ^2 (\theta _W)=0.2260 \pm 0.0048=> g^2/{4\pi}=0.0323$. It is hoped that this conflict between theory and measurement will be resolved when the supersymmetry or string theories take on a more complete form. Clearly present theory in general has some way to go but here we only wish to show that the characteristic value from our group $G_{29}$ representation theory for the value of $g^2/{4\pi}$ is no worse than the currently accepted theory suggested values, a theory often described as spectacularly successful in predicting the value of $\theta _W$. The value we obtain
ed for the fine structure constant is extraordinarily accurate in relation to the measured value. If there is an error in value this, it is of the order of one part in $10^{12}$ parts. Clearly this is not the case with the quantity $g^2/{4\pi}$. However, it is generally considered that the value for the fine structure constant that has been obtained from measurement is the {\it low\/} energy limit of a quantity that depends on the strength of energy momentum transfer involved in the coupling process. From the way the values of $\alpha$ and $g^2/{4\pi}$ are related by the present theory and the way the actual numerical values have been obtained it seems very likely that the value of $g^2/{4\pi}$, obtained by the present theory, is also definitely the low energy momentum transfer value. Thus, it is possible that the difference between theory and practice that seems to obtain here is just this difference between low energy theory structure and the higher energy range involved in meas
urement. Better agreement between theory and practice can be obtained by introducing {\it running\/} coupling constant factors into our definitions. However, here we shall be content with taking all the quantities involved as being definitely low energy representations.           
              
It seems that this type of cyclical system, with the many subgroup structures that can be called upon, is a potentially rich source for the numerical values of {\it fundamental\/} physical constants.
\vskip .5cm
\centerline{\bf 2 Kinematic Aspect}
\vskip .5cm 
The result of the previous section is essentially a theorem concerning relationships between polygons in the usual {\it static\/} Euclidian geometry. In this section, we examine geometrical and kinematic features of the movement of a physical object in a circular motion but with a quantized character. Quantized here, will mean the same as previously defined. We shall use the same symbols for quantities that arise in this context as were used in the previous section but initially assume no connection or equivalence of value with those quantities. Finally, the equality of equivalently denoted quantities will become apparent when we find that the kinematics of this section can be married with the static geometry of the previous section.

In Schr\"odinger quantum theory there is a well known result that the velocities of electrons in the first Bohr orbits of hydrogen like atoms assume an apparent quantized form. The speed $v_{B,Z}$ of an electron in the first Bohr orbit of the hydrogen like atom with the positive charge $Ze$ on its nucleus has the value $v_{B,Z}=Z\alpha c$ where $c$ is the speed of light and $\alpha$ is the fine structure constant. This result will be used as
a guide as to how quantized velocity in orbit can be defined in the context of orbital motion generally. It should be remarked that in quantum theory speed or velocity are not dynamical variables so that the orthodox view is that the quantization of velocity appearing in quantum theory is not usually credited with significance. The thinking here is at variance with this orthodox view.

Let us consider a physical object moving on a circular orbit. For such a configuration to occur there must be some force directed to wards a centre acting on the object. Other reasons for circular orbits are possible but here we consider the most common situation of a force directed towards a fixed centre. Let us assume that the speed of the object on the orbit is constant and its value is denoted by $v$ while its distance from the orbit centre is denoted by $r$. We shall also assume that we do not know the values of these two parameter and we do not know any relation between them other than the usual classical differential relation between the spatial position vector and the velocity. We shall accept the fundamental relativistic tenet that the speed in orbit is less than the speed of light $c$. Suppose that there is a naturally occurring quantum of speed and that its value is $\alpha c$. That is to say this quantum of speed $v_Q$ is obtained from the speed of light by multiplicat
ion by a fixed valued quantity $\alpha$. At this stage we shall consider that $\alpha$ is a numerical quantity of  {\it unknown\/} valued. All that we can say about the multiplier $\alpha$ at this stage is that its value must be less than unity if relativity is not to be violated,
$$v_Q=\alpha c,$$
where
$\alpha <1$ and $\alpha$ is an otherwise unknown quantity.
Suppose that $n_{max}$ is the largest {\it integer\/} for which $n_{max}\alpha c$
is still less than $c$, then
$$ n_{max}\alpha <1.$$
$v_c=v_{max}=n_{max} v_Q$ will thus be the largest {\it quantized\/} velocity that can occur in this circular orbit context.
Denote the radius of the orbit that a particle moving with this maximum quantized velocity would occupy by $r_c$. We use the integer $n_{max}$ to define a radial quantum length $l_c$ by
$$n_{max}l_c=r_c$$
Thus the circumference of the orbit $C_c$ will be given by
$$ C_c=2\pi r_c=2 n_{max}(l_c \pi)$$
Thus a natural circumferential length quantum $\pi l_c$ is induced with its own quantum number $2n_{max}$. The circumferential length quantum $\pi l_c$ divides the total circumference length up into $2n_{max}$ segments and so naturally induces the angular quantum
$$\chi _{n_{max}}= 2\pi /2n_{max}=\pi /n_{max}.$$
Returning to the relation $ n_{max}\alpha <1$ we see that we can define an angle $\chi ^*$ by putting
$$\cos (\chi ^*) = n_{max}\alpha$$
which simply uses the fact that the cosine function is always numerically less than unity. We define an outer radius $r_0$ by
$$r_0\cos (\chi ^*)=r_c.$$
 The radius $r_0$ is necessarily greater than the orbit radius $r_c$. We now give the angle $\chi ^*$ a definite location by choosing a special direction which can be any direction but is conveniently taken it to be the usual y-coordinate axis through the orbit centre. A second radial line through the orbit centre making the angle $\cos (\chi ^*)$ clockwise with the y-axis and terminating on the outer radius circle can then be used as the projection outer radius. There is also available the quantized angle $\chi _{n_{max}}$ which can be used to project an outer circle radius onto the special direction to generate another length along the special direction which will be called $r_b$.
$$r_b=\cos (\chi _{n_{max}})r_0$$
The outer circle being used here remains the same but the outer circle radius used for this projection is not necessarily the same outer circle radius as used for the $\chi ^*$ projection because in general $\chi _{n_{max}}$ is not equal to $\chi ^*$. Equality between these two angles would be a special case of some importance.  
All that has so far been done in this section is to set up notation expressing relations between the denoted quantities without any hint of the values quantities might have except that the upper limiting value $c$ of the velocity of light has been used.
Thus, collecting results, the unknown valued quantization multiplier $\alpha$ can be expressed in the form
$$\alpha = \cos (\chi ^*)/n_{max}.$$
The orbit radius can be expressed in the two ways
$$r_c=n_{max}l_c= r_0\cos (\chi ^*),$$
the velocity in orbit can be expressed in the the two ways
$$v_c=n_{max}\alpha c=\cos (\chi ^*)c$$
and we have defined the radial length $r_b$ by
$$r_b=\cos (\chi _{n_{max}})r_0.$$

Comparison of this list of relations with the relations in the first section, we see that connection can be made with the first section relations if $r_o$ and the quantum length $l_c$ are identified with either $r_{n_1,d}$ and the quantum length $l_{n_1}$
or $ r_{n_2,e}$  and the quantum length $l_{n_2}$ respectively with similarly subscripted quantities also identified. This choice is effectively the same as choosing between the two groups of the first section. Thus $n_{max}$ is to be either identified with $n_1$ or with $n_2$. From the kinematic relations of this section it follows that
$$r_c/r_0=v_{max}/c=v_c/c.$$
The meaning of this equality can easily be discovered. Consider a radial line from the origin through and moving with the object on the circle $C_c$ of radius $r_c$ with the speed $v_c=v_{max}$. If this moving radial line is extended to cut the outer circle of radius $r_0$ the intersection point so formed will have the velocity of light tangentially to the outer circle. Thus the choice of which group is to have the radius  $r_o$ associated with it also determines which group is to be associated with the limiting light velocity $c$. There is clearly a subtle symmetry operative in this context between the two groups and it is difficult to see how it can be expressed other than by the geometrical set up of the first section using the $n_1n_2$ order symmetry group with the two $n_1$ and $n_2$ order subgroup structure.
Of course, that is not to say there are not other ways of incorporating some such symmetry. Let us for the moment still not take on board the numerical consequences that stem from the properties of the $n_1n_2$ polygon representation for the group $G_{n_1n_2}$. Inspecting our collection of relations and definitions this section, we see that given numerical value for $\chi ^*$ all else can be determined in terms of the quantum number $n_{max}$. Thus if the angle $\chi ^*$ could be expressed in terms of the angle $\chi _{n_{max}}$ which is numerically known in terms of $n_{max}$ then the theory would be complete. The one or possibly two quantum numbers are to be regarded as input integers that define the system under discussion. There is a simple relation between the unknown $\chi ^*_n$ angles and the definite value $\chi _n$ angles,
$$ \cos (\chi ^* _{n_1})/ \cos (\chi ^* _{n_2})= \cos (\chi  _{n_1})/\cos (\chi  _{n_2}).$$
Or alternatively expressed,
$$ \cos (\chi ^* _{n_1})/ \cos (\chi  _{n_1})= \cos (\chi ^* _{n_2})/\cos (\chi  _{n_2}) =\beta (n_1n_2) =n_1n_2\tan (\chi _{n_1n_2})/\pi.$$
The last equality only holds if we accept the numerical consequences of the polygonal structure of the first section.
This last result also defines the useful function $\beta (n_1n_2)$ by the first equality above.
We also note that the set up of the first section implies that the quantum lengths $$l_{n_1}= r_{n_1,d}n_2\cos (\chi _{n_1})\tan (\pi/(n_1n_2))/\pi= r_{n_1,d} \cos (\chi _{n_1})\beta (n_1n_2)/n_1 $$ and $$l_{n_2}=r_{n_2,e} n_1\cos (\chi _{n_2})\tan (\pi/(n_1n_2))/\pi= r_{n_2,e} \cos (\chi _{n_2})\beta (n_1n_2)/n_2 $$
are naturally decomposable using the common sub-quantum length given by
$$l_{n_1n_2}= r_{n_1n_2,b}\tan (\pi/(n_1n_2))/\pi$$ $$= r_{n_1,d} \cos (\chi _{n_1})\beta (n_1n_2)/(n_1n_2)
= r_{n_2,e} \cos (\chi _{n_2})\beta (n_1n_2)/(n_1n_2).$$
$l_{n_1}=n_2l_{n_1n_2}$ and $l_{n_2}=n_1l_{n_1n_2}$.
These relations then imply in analogy with the first two characteristics the two smaller characteristics defined by
$$\alpha _{n_1,d} (n_1n_2)= l_{n_1n_2}/ r_{n_1,d}=\cos (\chi _{n_1})\beta (n_1n_2)/(n_1n_2)= \alpha _{n_1} (n_1n_2)/n_2$$
and
$$\alpha _{n_1,e} (n_1n_2)= l_{n_1n_2}/ r_{n_2,e}=\cos (\chi _{n_2})\beta (n_1n_2)/(n_1n_2)= \alpha _{n_2} (n_1n_2)/n_1.$$
These two characteristics are in the simple ratio
$$\alpha _{n_1,d}/\alpha _{n_1,e}=\cos (\chi _{n_1})/\cos (\chi _{n_2})  =\cos (\pi /137)/\cos (\pi /29)=1.005632147233...$$
The last equality being the numerical value of this ratio in the special case $n_1=137$ and $n_2=29$. Finally, we note from the earlier work a formula that generalises the Weinberg angle as another characteristic relation between representatives of the groups $G_{n_1}$ and $G_{n_2}$.
Calling the generalisation of the angle $\theta _W$, 
$\theta _G$ we have,
$$\sin ^2 (\theta _G)= \alpha_{n_1}(n_1\times n_2)/\alpha _{n_2}(n_1\times n_2).$$
For the case discussed earlier when $n_1=137$ and $n_2=29$ we can predict the low energy value for the Weinberg weak mixing angle $\theta _W=\theta _G$ as being given by
$$\sin ^2 (\theta _G)= \alpha_{137}(137\times 29)/\alpha _{29}(137\times 29)= 0.212871038465...$$
to twelve places of decimals.
\vskip .5cm 
\centerline{\bf 3 Conclusions}
\vskip .5cm 
It is a substantial step to have obtained $g$ and $\theta _W$ from the very simple basic cyclical group theory structure of the first section. From $g$ or $\theta _W$, using the value given by the relation $g^2/{4\pi}=\alpha _{29}(29\times 137)$, we can make predictions for the values of the other electro-weak coupling quantities $g^\prime=-e/\cos (\theta _W)$, $g_Z= g^\prime /2\sqrt 2$ and $\alpha _W= (g^2/4\pi )/8$. Further, the mass ratio of the Z and W gauge bosons is given by
$$M_Z/M_W=(1+(g^\prime /g)^2)^{1/2}=1/\cos (\theta _W)=1/0.887$$ and, as with the rest of these quantities, the physical measured numerical value is not known with anything like the accuracy that the fine structure constant value is known from measurement. The experimental value of the mass ratio $M_Z/M_W=1/0.881$. 
\vskip .5cm 
\centerline{\bf 4 References}
\vskip .5cm 
\leftline{1 Georgi H. and Glassow S. L., Phys. Rev. Lett. {\bf 28}, 1494 (1972)}
\leftline{2 Weinberg S, The Theory of Quantum Fields, {\bf 2}, Cambridge UP. (1996)}
\leftline{3 Gilson J. G. Calculating the Fine Structure Constant, Phys. Essays. {\bf 9}, No. 2, 342 (1996)}
\leftline{4 Gilson J. G. Relativistic Length Contraction and Quantization, PIRT Conf. 1998}
\leftline{5 Martin B.R. \& Shaw, G., Particle Physics (Second Edition), John Wiley \& Sons Ltd.(1997)}
\leftline{6 Gilson J. G. Anouncements, The Times Newspaper 21st July 1999}
\leftline{7 Gilson J. G. Finite Renormalization II\footnote{*}{Updated version of reference 8}, Late Papers PIRT Conf. 1998}
\leftline{8 Gilson J. G. Finite Renormalization, Speculations in Science and Technology,}
\leftline{{\bf 21}, No 3, September 1998}

\end